# Turbulence and Fossil Turbulence in Oceans and Lakes

**PakTao Leung**

Department of Mechanical and Aerospace Engineering,
University of California San Diego, La Jolla CA 92093-0411, USA
ptleung@ucsd.edu, **http://mae.ucsd.edu/~ptleung/**

**Carl H. Gibson**

Departments of Mechanical and Aerospace Engineering
and Scripps Institution of Oceanography,
University of California San Diego, La Jolla CA 92093-0411, USA
cgibson@ucsd.edu, **http://www-acs.ucsd.edu/~ir118/**

## ABSTRACT

Turbulence is defined as an eddy-like state of fluid motion where the inertial-vortex forces of the eddies are larger than any of the other forces that tend to damp the eddies out. Energy cascades of irrotational flows from large scales to small are non-turbulent, even if they supply energy to turbulence. Turbulent flows are rotational and cascade from small scales to large, with feedback. Viscous forces limit the smallest turbulent eddy size to the Kolmogorov scale. In stratified fluids, buoyancy forces limit large vertical overturns to the Ozmidov scale and convert the largest turbulent eddies into a unique class of saturated, non-propagating, internal waves, termed fossil-vorticity-turbulence. These waves have the same energy but different properties and spectral forms than the original turbulence patch. The Gibson (1980, 1986) theory of fossil turbulence applies universal similarity theories of turbulence and turbulent mixing to the vertical evolution of an isolated patch of turbulence in a stratified fluid as its growth is constrained and fossilized by buoyancy forces. Quantitative hydrodynamic-phase-diagrams (HPDs) from the theory are used to classify microstructure patches according to their hydrodynamic states. When analyzed in HDP space, previously published oceanic data sets show their dominant microstructure patches are fossilized at large scales in all layers. Laboratory and field measurements suggest phytoplankton species with different swimming abilities adjust their growth strategies by pattern recognition of turbulence-fossil-turbulence dissipation and persistence times that predict survival-relevant surface layer sea changes. New data collected near a Honolulu waste-water outfall show the small-to-large evolution of oceanic turbulence microstructure from active to fossil states, and reveal the ability of fossil-density-turbulence patches to absorb, and vertically radiate, internal wave energy, information, and enhanced turbulent-mixing-rates toward the sea surface so that the submerged waste-field can be detected from a space satellite.

# 1. INTRODUCTION

What are the fundamental properties of turbulent flows? What is the definition of turbulence? What is the direction of the turbulence cascade? What happens if the ambient density of an ocean or lake is stably stratified? Do random fluctuations of temperature in a patch of microstructure prove that the fluid is turbulent? Widespread disagreement about the answers to these questions has greatly complicated the study of oceanic and lake turbulence. The subject has become somewhat controversial so the reader should be warned that the views that follow are those of the authors and may not reflect those of everyone else (or anyone else) in the community. For a range of different and sometimes similar views about stratified turbulence see DeBruynKops and Riley (2003), Fernando (1988), Gourlay et al. (2001), Itsweire et al. (1993), Ivey and Imberger (1991), Ivey et al. (1992), Luketina and Imberger (1989), Smyth and Moum (2000), Winters and D'Asaro (1996), Winters et al. (1995), Fan Zhisong (2002) and other references discussed below in context.

Turbulence is a property of fluid flows that has been notoriously difficult to define. Oceanographers such as Gregg (1987) and Fan Zhisong (2002) treat turbulence as indefinable with classes of turbulence existing in the ocean that have no counterpart in the laboratory. Libby (1996) requires a wide spectrum of velocity fluctuations to distinguish turbulence from "unsteady laminar flow". Frisch (1995) also emphasizes the importance of high Reynolds numbers. Stewart (1969) offers a syndrome definition that lists various commonly accepted properties of turbulence. Similarly, Tennekes and Lumley (1972) lists irregularity, diffusivity, large Reynolds number, three-dimensional vorticity fluctuations, and large dissipation rates as distinguishing properties of turbulence, and this approach is continued by Pope (2000). Unfortunately such broad syndrome-definitions give the impression that turbulence has no intrinsic scientific basis and may be freely defined according to the needs of the situation. A more precise, narrow, definition is necessary, especially for stratified and rotating turbulence applications, and is provided implicitly by Gibson (1980) and explicitly by Gibson (1991a, 1996, 1999). The turbulence definition, based on the inertial-vortex force $\vec{v} \times \vec{\omega}$, is designed to exclude all flows and mixing processes as non-turbulent that depart from universal similarity laws of Kolmogorov and the extension of these laws to scalar fields mixed by turbulence, Gibson (1991a), even though they might fit turbulent-syndrome-definitions frequently used in oceanography. Fossil turbulence is an example, from the Woods (1969) fossil turbulence workshop papers. Random fluctuations in hydrophysical fields of the ocean must have supercritical Reynolds, Froude and Rossby numbers to be turbulence according to the narrow turbulence definition recommended here.

Universal Kolmogorov similarity for wind tunnel, water tunnel and ocean tidal channel flow turbulence and universal Batchelor (1959) scalar similarity for the high Prandtl number oceanic scalar fields temperature (Pr ≈ 10) and salinity (Pr ≈ 1000), were first demonstrated by Gibson and Schwarz (1963). These concepts are now generally accepted among oceanographers. The "Nasmyth spectrum" (Nasmyth 1970) further confirms that Kolmogorov similarity occurs in the ocean, and Oakey (1982) confirms that the universal Batchelor spectral form applies to turbulent temperature in the ocean.

Our physically-motivated definition of turbulence follows from the momentum conservation equations given in the next section. The nonlinear term $(\vec{v} \cdot \nabla)\vec{v}$ of the Navier-Stokes equation is decomposed into the gradient of the kinetic energy per unit mass $\nabla(v^2/2)$ so that $v^2/2$ can be included in the (often nearly constant) Bernoulli group B of mechanical energy terms (B = kinetic $v^2/2$ + enthalpy $p/\rho$ + potential $gz$), and a lift-acceleration term $\vec{v} \times \vec{\omega}$ that we call the inertial-vortex-force per unit mass. This unique nonlinear term is responsible for turbulence and its dynamics and therefore deserves to be included explicitly in any definition of turbulence. Since $\vec{v} \times \vec{\omega}$ is zero for irrotational flows, such flows are non-turbulent by definition. The commonly accepted notion that turbulence cascades from large scales to small arises because large-scale irrotational flows often supply energy to smaller-scale turbulent flows. When $\vec{v} \times \vec{\omega}$ is included in the turbulence definition the direction of the turbulent energy cascade from small to large is emphasized. All turbulent flows form at small scales and cascade to larger scales through a self-similar eddy-pairing mechanism driven by $\vec{v} \times \vec{\omega}$ forces at larger and larger scales. This is the physical mechanism and basis of the 1941 Kolmogorov universal similarity theory of turbulence and the Planck-Kerr instability of the big bang, Gibson (2003ab), where Planck scale inertial-vortex forces balance quantum gravity to produce space, time, energy, entropy and the universe. The Gibson (1980) theory of fossil temperature, salinity, and vorticity turbulence in the stratified ocean is based on the application of universal similarity theory to the growth of a small, powerful, patch of turbulence that grows to larger scales where it is fossilized by buoyancy forces. The signature of fossil-turbulence patches would not be unique if turbulence could originate at large scales as nearly inviscid Kelvin-Helmholtz billows as assumed by Gregg (1987), Smyth et al. (2001), Fan Zhisong (2002) and others.

Because broad definitions of turbulence are employed in the interpretation of most oceanic and limnological turbulence studies, the reliability of the resulting conclusions becomes questionable. Without our proposed narrow definition of turbulence, the unique signature of fossil turbulence is lost and all scalar and vector microstructure is lumped together as turbulence. Data sets that contain fossilized turbulence patches are mistaken as representative of the turbulence process, even though few or none of the patches may actually be turbulent at the time of sampling, Finnigan et al. (2002). The tendency is to vastly underestimate the vertical turbulent fluxes of heat, mass and momentum according to Gibson (1987, 1991abcd). From the available data we suggest fossil turbulence effects must always be taken into account in the analysis of oceanic and limnological stratified microstructure data. No data set is complete unless fully active patches of the dominant turbulent events are included or inferred from fossil turbulence theory. The proposition that turbulence exists anywhere vorticity or mixing are above arbitrary thresholds, so that data sets with many microstructure patches are by definition representative, as put forth by Gregg (1987), is accepted as a working hypothesis by a significant fraction of the oceanographic microstructure community, but is theoretically incorrect and potentially quite misleading in practical applications, Gibson (1987, 1999). It is claimed by Smyth et al. (2001) that inviscid, non-turbulent,

Kelvin-Helmholtz billows are the large scale source of stratified turbulence in the ocean based on very low Reynolds number direct numerical simulations of stratified turbulence that show a monotonic increase of the Ozmidov to Thorpe displacement scale ratio (see §3.2), consistent with the Gregg (1987) model. Large "pre-turbulence" billows form, collapse by gravity to form turbulence, and vanish without a trace, so that data sets containing any microstructure can be considered representative. This phenomenology of Smyth et al. (2001), Gregg (1987) and Fan Zhisong (2002) is completely orthogonal to that presented here and is contrary to oceanic measurements discussed in §4. We suggest turbulence always starts with large dissipation rates and small overturning scales and grows to persistent large-Thorpe-scale small-ε fossils, giving a monotonic decrease in the Ozmidov/Thorpe scale ratio, and leaving highly persistent fossil-turbulence remnants in all hydrophysical fields. According to our paradigm of stratified oceanic and lake turbulence, it is mandatory to examine the hydrodynamic state of the dominant patches of microstructure using calibrated hydrodynamic phase diagrams to avoid undersampling. Our recommended laboratory HDP calibrations are described in Gibson (1987).

This paper describes methods to identify from microstructure measurements different stages of turbulence as it evolves from completely-active to active-fossil to completely-fossilized hydrodynamic states in a stratified fluid. Section §2 discusses turbulence theory, since most misconceptions about fossil turbulence result from misconceptions about turbulence. Section §3 defines fossil turbulence and shows how it can be identified using hydrodynamic phase diagrams (HPDs). Section 4 discusses fossil turbulence measured in the laboratory and ocean using a summary HPD from Gibson (1996), and shows a recent application of fossil turbulence theory and HPDs to interpret sea-truth microstructure measurements for the remote sensing of submerged fossil turbulence. Information about the location of a submerged municipal waste field trapped by buoyancy was found to propagate vertically to the sea surface and was detected from a space satellite image. Two other oceanic applications of fossil turbulence are discussed: the resolution of the dark mixing paradox, and evidence that swimming and non-swimming phytoplankton growth rates respond to turbulence and fossil turbulence in surface layers of the sea. Finally, a summary and conclusions are provided in §5.

## 2. TURBULENCE THEORY

This section addresses two common misconceptions of oceanic turbulence. One is the idea that random microstructure in velocity, temperature, or any other hydro-physical field is sufficient evidence of turbulence ("anything that wiggles is turbulence"). The other is the idea that turbulence is formed first at large scales by inviscid Kelvin-Helmholtz billows that later form smaller scales by gravitational collapse. Both ideas are defended at length by Gregg (1987), by others in the oceanographic microstructure community as listed in his invited review article, and by Smyth et al. (2001) from numerical studies. However, Caulfield and Peltier (2000) conclude from a different numerical study that inertial forces dominate buoyancy forces in the formation of turbulence. In defense of the oceanographers it should be noted that the misconception about the direction of the turbulence cascade has infested the turbulence and fluid mechanics literature for decades, starting with the Richardson (1922) doggerel "Big whorls have little whorls, that feed on their velocity, and smaller whorls have smaller whorls, and so on to viscosity---in the molecular sense", and persists today (see Gibson 1999 for a revised version of the poem[1]). Fossil turbulence cannot be distinguished from turbulence if either of these ideas is true.

The term "turbulence" as defined in oceanography has evolved to become progressively narrower over time. Sixty years ago, Sverdrup et al. (1942) asserted that most of the ocean is turbulent, but today most oceanographers believe that at any instant of time only a small fraction (about 5%) of the ocean is fully turbulent.[2] The reason for the shift in estimated turbulence fraction is mostly due to a shift in the accepted definition of turbulence among oceanographers.

## 2.1 Definition of turbulence

We define turbulence based on the nonlinear inertial-vortex force term $\vec{v} \times \vec{\omega}$ in the momentum conservation equations written in the form

$$\frac{\partial \vec{v}}{\partial t} = \vec{v} \times \vec{\omega} + \vec{v} \times 2\vec{\Omega} - \nabla B + \nabla \cdot (\vec{\tau}/\rho) + \vec{b} + ... \qquad (1)$$

where $\vec{v}$ is the velocity, t is time, $\vec{\omega} = \nabla \times \vec{v}$ is the vorticity, $\vec{\tau}$ is the viscous stress tensor, and the mean density $\rho$ is assumed constant. The Bernoulli group of mechanical energy terms $B = v^2/2 + p/\rho + gx_3$, where p is the pressure, g is gravity, and $x_3$ is up, is constant for steady, irrotational, inviscid flows (the Bernoulli equation). The buoyancy force $\vec{b}$ depends on a vertical gradient in the mean density. The Coriolis force is $\vec{v} \times 2\vec{\Omega}$, where $\vec{\Omega}$ is the angular velocity of the Earth. Electromagnetic, surface tension, and other forces are omitted for simplicity. Turbulence occurs when the inertial-vortex forces, $\vec{v} \times \vec{\omega}$, exceed the viscous forces, $\nabla \cdot (\vec{\tau}/\rho)$, if other forces are negligible. The ratio of inertial-vortex forces $\vec{v} \times \vec{\omega}$ to viscous forces is the Reynolds number, $Re = UL/\nu$, where $U$ is a characteristic velocity and $L$ is a characteristic length scale of a flow. "Characteristic" means locally or globally averaged, depending on the flow of interest.

Similar dimensionless groups arise from the other terms in (1). All of these must also be smaller than $\vec{v} \times \vec{\omega}$ for the flow to be turbulent, by definition (Table 1).

---

[1] "Little whorls on vortex sheets, form and pair with more of, whorls that grow by vortex forces: Slava Kolmogorov!"

[2] About 5% of upper oceanic interior layers have been turbulent recently based on the volume fraction of temperature microstructure typically encountered by towed microstructure sensors (Washburn and Gibson 1982, 1984). Most of the microstructure patches are fossilized, with less than 5% classified as "actively turbulent" using hydrodynamic phase diagrams discussed below, giving about 0.3% turbulence. The oceanic fossil-turbulence to active-turbulence fraction approximately equals the Reynolds number ratio $\varepsilon_0/\varepsilon_F$ of the dominant turbulence events of a particular layer (§3.2), which varies from about $10^3$ in surface layers to $10^4$ or more in abyssal layers, with a maximum of about $10^7$ near the equator.

| Richardson Number (sub-critical to be turbulent) | the squared ratio of $\vec{b}$ to inertial-vortex forces ($\vec{v} \times \vec{\omega}$) | $Ri \equiv \dfrac{N^2}{(\partial U / \partial z)^2}$ | $N$ is Väisälä frequency $\rho$ is density g is gravity $U$ is the velocity in the horizontal, $z$ is down |
| Froude number (super-critical to be turbulent) | the ratio of $\vec{v} \times \vec{\omega}$ to $\vec{b}$ (normally to 1/2 power) | $Fr \equiv 1 / Ri^{1/2}$ $= U / LN$ | $U$ is a characteristic velocity $L$ is a characteristic length scale of the flow. |
| Rossby number (super-critical to be turbulent) | the ratio of $\vec{v} \times \vec{\omega}$ to $\vec{v} \times 2\vec{\Omega}$ | $Ro \equiv \dfrac{\vec{v} \times \vec{\omega}}{\vec{v} \times 2\vec{\Omega}}$ | $\vec{v} \times 2\vec{\Omega}$ is the Coriolis Force |
| Reynolds number (super-critical to be turbulent) | The ratio of $\vec{v} \times \vec{\omega}$ to viscous forces | $\mathrm{Re} = \dfrac{UL}{\nu}$ | $U$ is a characteristic velocity $L$ is a characteristic length scale of the flow. |

***Table 1***: Dimensionless groups formed from the ratio between inertial-vortex forces ($\vec{v} \times \vec{\omega}$) and other terms in the momentum conservation equations.

It is commonly known that critical Reynolds, Froude, and Rossby numbers must be exceeded to permit turbulence as shown in Table 1. Other forces like surface tension or electromagnetic forces can also inhibit turbulence and have corresponding critical dimensionless numbers for turbulence. The inertial-vortex force causes vortex sheets to be unstable because it is perpendicular to the vortex sheet, and is such that any perturbation of the sheet will be amplified in the direction of the perturbation to form eddies. This is the basis of our definition of active turbulence.

**Definition:** *Turbulence is an eddy-like state of fluid motion where the inertial-vortex forces of the eddies are larger than any of the other forces which tend to damp them out.*

This definition, discussed in Gibson (1991a, 1996, 1999), accommodates both three-dimensional and two-dimensional turbulence. Irrotational flows with zero vorticity are non-turbulent by our definition because the inertial-vortex forces per unit mass $\vec{v} \times \vec{\omega}$ are zero. Furthermore, flows like internal waves that are dominated by buoyancy forces are non-turbulent by definition, even though the waves are rotational, non-linear, random, and may even be caused by turbulence. Coriolis-inertial waves formed when two-dimensional turbulence is damped by Coriolis forces can be eddy-like, non-linear, and random. However, such waves are not turbulent by definition because the Coriolis forces dominate the inertial-vortex forces. Eddy-like flows in cavities of boundary layers are random, rotational, and diffusive, but are not turbulent by definition because the eddies are dominated by viscous forces.

**2.2 Formation of turbulence**

The following physical model illustrates the significant role of inertial-vortex forces $\vec{v} \times \vec{\omega}$ in the formation of turbulence. Any flow that produces a shear layer is likely to produce turbulence. The reason is that perturbations of the shear layer induce $\vec{v} \times \vec{\omega}$ forces in the direction of the perturbation, so that the perturbation grows. Such positive perturbation feedback is what is meant by the term "instability". Shear layers are produced by 1) flow around obstacles, 2) jets into stagnant bodies of fluid, 3) boundary layers, or 4) mixing layers. Figure 1 shows a shear layer (mixing layer) and the evolution of velocity perturbations to eddy-like motions because they induce inertial-vortex forces in the perturbation direction. The domination of $\vec{v} \times \vec{\omega}$ forces first causes the shear layer to wiggle and then curl up and form the first turbulent eddy. Turbulence arises because of this shear instability which is induced by the growth of inertial-vortex forces.

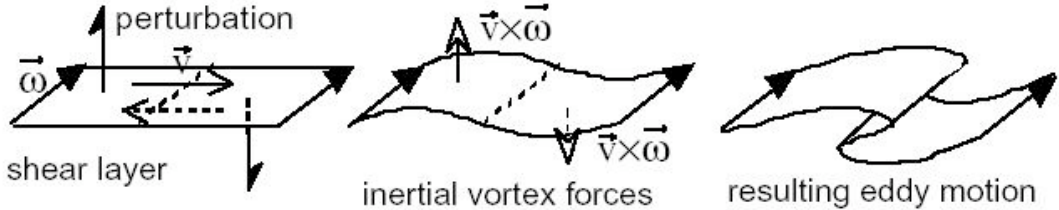

**Figure 1.** Fundamental mechanism of turbulence: perturbations on a shear layer cause inertial-vortex $\vec{v} \times \vec{\omega}$ forces in the two opposite perturbation directions, with consequent eddy formation. Thin shear layers (vortex sheets) first thicken by viscous diffusion to the Kolmogorov scale before forming eddies.

Shear layers like that shown in Figure 1 are unstable to perpendicular velocity perturbations $\delta v(L)$ at all scales L. However, eddies that form first are those with the smallest overturn time T(L) = L/$\delta v(L)$, since usually $\delta v(L)$ is weakly dependent on L such as for turbulence with $\delta v(L) \propto L^{1/3}$ so that T(L) $\propto L^{2/3}$ decreases with L from the Kolmogorov (1941) second universal similarity hypothesis. Viscosity limits the smallest eddy size to the Kolmogorov scale

$$L_K = \left( v^3 / \varepsilon \right)^{1/4} \tag{2}$$

where $v$ is the kinematics viscosity of the fluid and $\varepsilon$ is the viscous dissipation rate

$$\varepsilon = 2 v e_{ij} e_{ij} \tag{3}$$

$e_{ij}$ is the rate of strain tensor,

$$e_{ij} = \frac{1}{2} (\partial v_{i,j} + \partial v_{j,i}) \tag{4}$$

$v_i$ is the i component of velocity, and $\partial v_{i,j}$ is the partial derivative of $v_i$ in the j direction, and repeated indices are summed over i and j = 1 to 3. It turns out that viscous forces prevent any eddies smaller than about $5L_K$, the scale of the universal critical Reynolds number. Such eddies have sub-critical Reynolds numbers.

## 2.3 The turbulence cascade

After formation, nearby eddies with the same average vorticity induce velocities such that they will orbit about a point between them. Hence, they pair up and form a larger eddy. This process will repeat until the nested pairs of paired eddies reach a length scale for which inertial-vortex forces no longer dominate. The net transfer of energy and momentum between eddies of comparable size from small to large scales is referred as the turbulence cascade. It starts from small scales and grow to larger scales, with continuous feed back from large to small. As shown in Figure 2 for an evolving shear layer, eddies draw energy from the irrotational fluid layers moving in opposite directions on opposite sides of the shear layer. This process brings in larger scale kinetic energy to increase the size and strength of the turbulence. A non-turbulent energy cascade in the external, irrotational fluid from large scales to small is induced by the turbulence, and should not be confused with the net turbulence energy cascade from small scales to large.

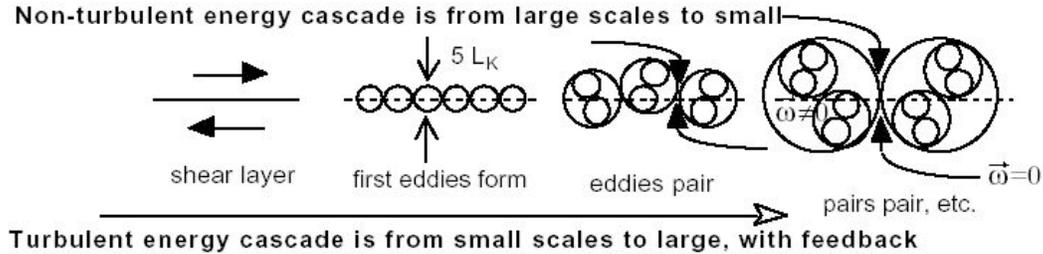

***Figure 2.*** Schematic of the turbulence cascade process, from small scales to large. Transition occurs at scale $Re_{crit}^{1/2} L_K$, where empirically the universal critical Reynolds number $Re_{crit}$ is about 25.

The irrotational fluid entrained at large scales by a turbulent flow is the source of mass, momentum, and kinetic energy for the turbulence, and is also the source of the common misconception that turbulence cascades from large scales to small. Turbulent flows such as jets, wakes, or boundary layers (Figures 2, 3 and 4) do not shrink but grow in what is wrongly termed an "inverse cascade". If the external flow is irrotational then its inertial-vortex forces are zero and the energy cascade of the flow from large to small scales is non-turbulent, by definition.

To summarize, inertial-vortex forces induce a shear instability that produces the first turbulent eddy at the smallest possible scale permitted by viscosity, the Kolmogorov scale. The first eddies then pair with others and form larger eddies, extracting energy from the non-turbulent flow. The turbulent eddies grow in this fashion until they reach the maximum scale allowed by boundaries or the other forces. The maximum vertical size may be limited by buoyancy forces, and Coriolis forces may limit horizontal growth in stratified oceanic flows on the rotating earth. What happens to the turbulent eddies and the scalar field microstructure produced after the patches grow to maximum size? Will the partially mixed temperature, density and salinity microstructure patches collapse and disappear without a trace? Will the vorticity, density, temperature and salinity microstructure patches produced by turbulence persist as fossil turbulence? Will the fossil turbulence patches simply preserve information about the past turbulence event, or can they mimic turbulence and absorb ambient energy and dominate the vertical diffusion of ambient properties? The next section addresses these questions.

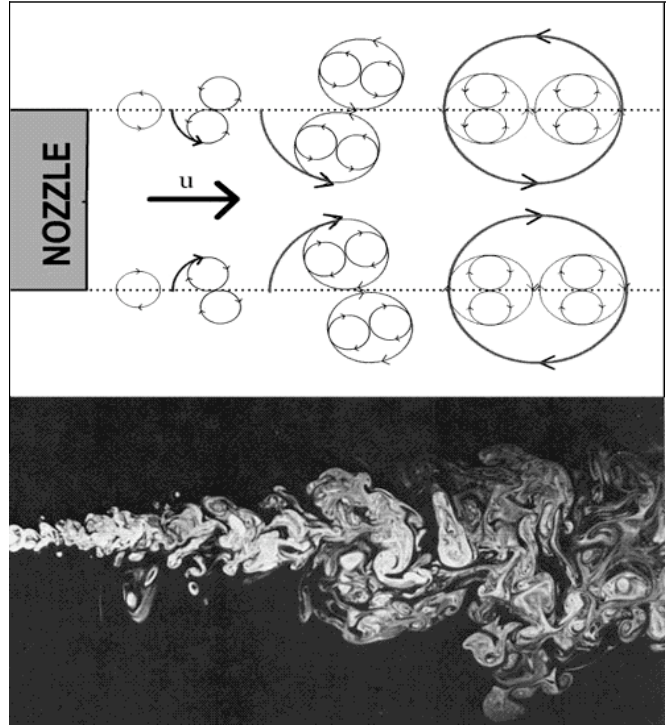

**Figure 3.** The growth of a nonstratified turbulent jet. Small eddies of turbulence form first, before the larger eddies downstream.. The figure on top is a schematic of the nozzle flow, while the bottom one is a picture taken from a laboratory jet (Van Dyke 1982, p97).

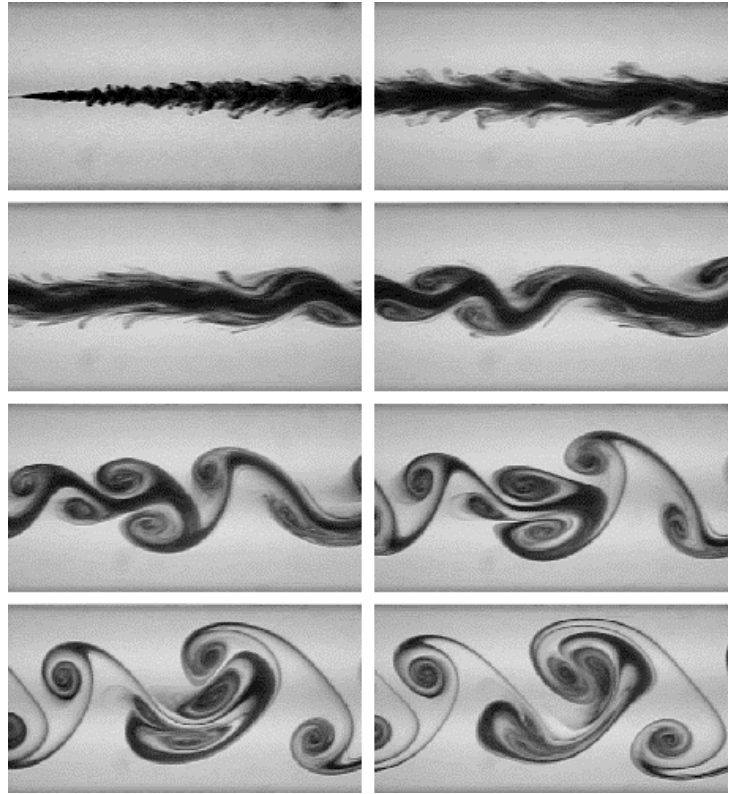

**Figure 4.** Top view of a stratified turbulent jet-wake (Smirnov and Voropayev 2001). Turbulent eddies start from small scales, then pair with another similar size eddy, and the pairs in turn pair with other pairs before fossilization. The duration of this sequence is 166 seconds, while the vertical size of each frame is 33 cm.

## 3.  FOSSIL TURBULENCE THEORY

A very important characteristic of turbulence is that it produces highly persistent, irreversible effects in a wide variety of hydro-physical fields. Linear waves come and go without leaving any trace, but turbulence is intrinsically irreversible and leaves remnant hydro-physical signatures, especially in natural flows where turbulence is damped out at its largest scales. Such signatures are described by the term "fossil turbulence". Like fossil dinosaurs are the evidence and proof of previously existing dinosaurs, fossil turbulence is the evidence and proof of previous turbulent events. Familiar examples include the contrails of jet aircraft in the stratified atmosphere. Buoyancy forces damp the large scale turbulence within a few airplane lengths, but the fossil turbulence contrails persist and preserve information about the turbulence that produced them. The turbulence process and turbulent mixing process is self similar.  Each stage of the turbulence cascade process is the same as every other stage.  This is the basis of universal similarity theories of turbulence and mixing and why it is impossible to tell the Reynolds number of a photograph of turbulence mixing a tracer such as smoke or fog if the Kolmogorov scale is not resolved.  This is why it works to use inexpensive models of burning ships or buildings in movies rather than the real thing.

**3.1 Definition of fossil turbulence**

Many hydro-physical fields are scrambled during an actively turbulent event. However, some "footprints" may persist independently for different periods of time. They preserve the event information as fossil turbulence long after the turbulence has died away. Assuming the definition of active turbulence given previously, fossil turbulence may be defined as follows,

**Definition:** *Fossil turbulence is a fluctuation in any hydro-physical field produced by turbulence that persists after the fluid is no longer actively turbulent at the scale of the fluctuation.*

Fossil turbulence appears in a variety of hydro-physical fields. The terminology fossil temperature turbulence indicates fluctuations of temperature produced by turbulence that persist in the fluid after it is no longer turbulent at the scale of the temperature microstructure. Fossil vorticity turbulence refers to the unique class of internal wave motions at the ambient stratification frequency N produced when buoyancy forces overcome the inertial vortex forces of the turbulence.

**3.2 Formation of fossil turbulence**

Gibson (1980) introduced a theory of fossil turbulence based on the evolution of a growing, powerful, isolated turbulence patch. The recommended experiment to observe the process is pouring cold milk into hot coffee in a large clear cup. It was assumed from observations that the turbulence of the patch starts at the smallest possible scale, the Kolmogorov scale (equation 2), and grows by an eddy pairing process to the largest possible scale. As mentioned above, other forces will become more significant as the turbulent eddies grow larger. When the eddies reach a certain critical scale, the density gradient they are trying to overturn is no longer passive but begins to overwhelm their inertial-vortex forces. The eddies then fail to overturn, fall back and start a bobbing motion that is not turbulence by definition but a unique class of internal waves (fossil vorticity turbulence). This happens when the vertical length scale at which the buoyancy force is of the same order of magnitude as the inertial forces, which is the Ozmidov scale

$$L_R \equiv \left(\varepsilon/N^3\right)^{1/2} \tag{5}$$

where the dissipation rate $\varepsilon$ has value $\varepsilon_o$, the value at beginning of fossilization, $\nu$ is the kinematics viscosity and N is the Väisäilä frequency ($N \equiv \sqrt{-g(\partial\rho/\partial z)/\rho}$) of the ambient fluid. The buoyancy dominated motions of the fossil vorticity turbulence persist for long periods.

The kinetic energy of the turbulence is converted to the internal wave energy of trapped fossil vorticity turbulence motions within the fossilized patch. When the turbulence reaches complete fossilization, its dissipation rate is

$$\varepsilon_F = 30\nu N^2 \tag{6}$$

as derived by Gibson (1980). Because $\varepsilon_F$ is much less than $\varepsilon_o$ and the kinetic energy of the fossil is the same as the kinetic energy of the turbulence patch just before fossilization, the persistence time is greatly increased if $\varepsilon_o/\varepsilon_F \gg 1$.

Only the largest scales of the fossil temperature turbulence spectrum are affected by the fossilization process. The small scales continue to dissipate and mix strongly from the small scale turbulence of the active-fossil HPD quadrant. From the Gibson (1980) theory, this provides a means of estimating the dissipation rate at the beginning of fossilization $\varepsilon_o$

$$\varepsilon_o \approx 13 D C_x N^2 \qquad (7)$$

where D is the molecular diffusivity of the scalar field like temperature, and $C_x \approx C/3$ is the streamwise Cox number. The Cox number C for temperature is the mean square temperature gradient over the square mean temperature gradient

$$C = \frac{\overline{(\nabla T)^2}}{(\nabla \overline{T})^2} \qquad (8)$$

The dissipation rate at the beginning of fossilization $\varepsilon_o$ can also be estimated from the maximum Thorpe overturn scale of the turbulent patch

$$\varepsilon_o \approx 3 L_{T\max}^2 N^3 \qquad (9)$$

using $L_{T\max} \approx \sqrt{3} L_{R_o}$, Gibson, Nabatov and Ozmidov (1993), where $L_{Tmax}$ is the maximum Thorpe displacement of the patch and N is the ambient Väisäilä frequency. The Thorpe density displacement is the vertical distance particles must be moved to achieve a monotonic vertical density profile from a density profile with overturns.

Because the dissipation rates at beginning and end of the fossilization are very different but the kinetic energies remain the same, the fossil turbulence patch will persist by a factor of $\varepsilon_o/\varepsilon_F$ longer than a turbulence patch with $\varepsilon = \varepsilon_o$ without stratification. This ratio may be $10^4$ to $10^6$ for the dominant patches of interior oceanic layers and only a small fraction of such layers contains any microstructure, either active or fossil. Many thousands or millions of fossilized dominant patches must therefore be sampled before an active dominant patch is discovered so that the data set may be considered representative, requiring either very long data records in the layer before reliable average dissipation rates can be estimated statistically, Baker and Gibson (1987), or estimated by fossil turbulence models. A contrary view expressed by Smyth et al. (2001) is widely accepted judging from the sparse sampling strategies commonly employed in interpreting vertical fluxes in deep ocean layers, and the lack of HPD interpretation of dominant patches. Suppose the dominant overturning patches for a deep layer of the ocean interior were 10 meters in vertical extent and occupy a fraction $10^{-3}$ of the layer with $\varepsilon_o/\varepsilon_F$ about $10^5$. Thus, $10^9$ m of independent data record for the layer must be examined for a 50% chance of encountering an actively turbulent dominant patch. At a velocity of 1 m/s this would require about 30 years using a single sensor. Since the sensor would encounter a fossil patch about once every three hours, the motivation for developing a reliable system of hydro-paleontology is clear.

### 3.3 Identification of fossil turbulence

Microstructure patches can easily be classified according to their hydrodynamic state using a hydrodynamic phase diagram, as shown schematically in Figure 5. The hydrodynamic phase diagram (HPD) concept was introduced by Gibson (1980) as a means of classifying microstructure patches according to their hydrodynamic states: active, active-fossil, or fossil turbulence depending on $\varepsilon$, $\varepsilon_o$ and $\varepsilon_F$. In the diagram, the Froude number of the patch Fr normalized by $Fr_o$ at the beginning of fossilization is plotted against the Reynolds number normalized by $Re_F$ of the patch at complete fossilization. Confirmation of the recommended universal constants is discussed in Gibson (1987, 1999).

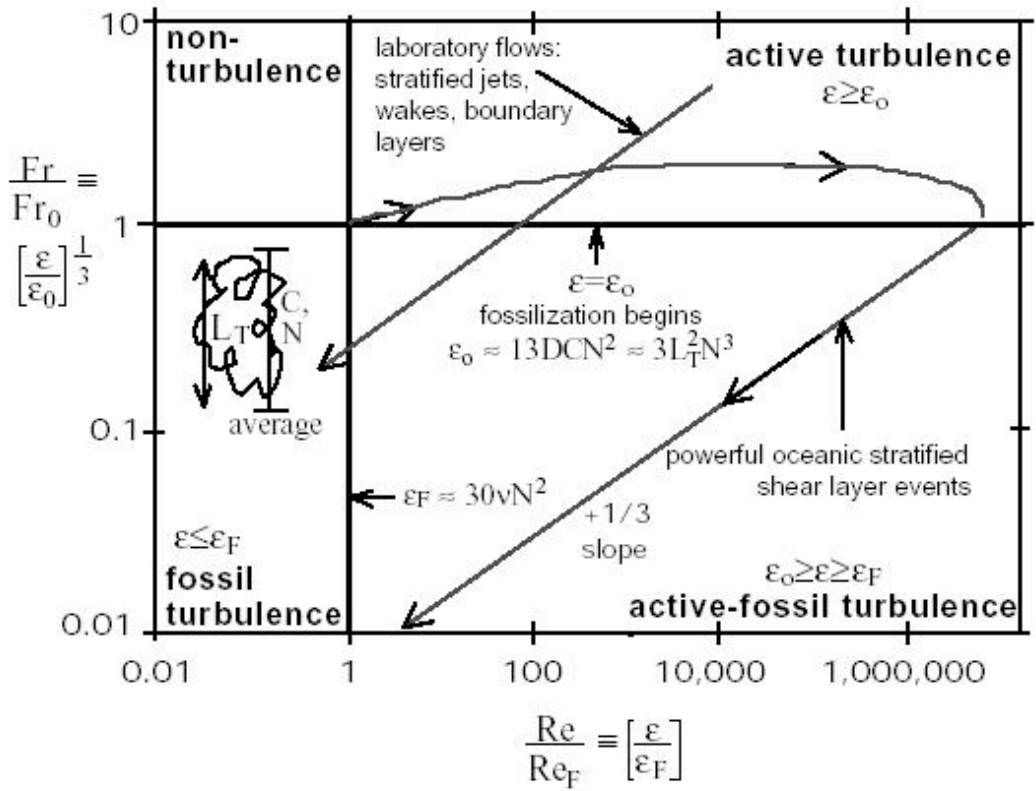

**Figure 5.** Hydrodynamic phase diagram (HPD) used to classify microstructure patches according to their hydrodynamic state in a stratified fluid. Trajectories for the evolution of stratified laboratory and dominant oceanic turbulent events are indicated by arrows and lines. Note that the dissipation rate $\varepsilon$ is measured for the patch, and the normalized Froude number and Reynolds numbers computed from the expressions for $\varepsilon_o$ and $\varepsilon_F$ given on the diagram. The maximum Thorpe overturn scale, $L_T$, and the Cox number, C, can be used to estimate $\varepsilon_o$. The ambient N for the patch can be used to find $\varepsilon_F$, where C and N must be averaged over vertical scales larger than the patch, as shown in the inserted sketch.

The ordinate of the HPD in Figure 5 is derived from the definition of Froude number in Table 1, $Fr = U/LN$, using $U \approx (\varepsilon L)^{1/3}$ for the characteristic velocity U of turbulence from Kolmogorov's second universal similarity hypothesis. The normalized Froude number $Fr/Fr_o = (\varepsilon/\varepsilon_o)^{1/3}$ is the ratio between the Froude numbers of the patch as detected with dissipation rate ε and a patch with the same L and N values at beginning of fossilization with dissipation rate $\varepsilon_o$.

The abscissa of the HPD in Figure 5 is derived from the Table 1 definition of Reynolds number $\mathrm{Re} = UL/\nu = c\varepsilon/\nu N^2$, using $U \approx (\varepsilon L)^{1/3}$ with L taken as the maximum turbulence scale in the patch $L = L_R \propto (\varepsilon/N^3)^{1/2}$, so that $\mathrm{Re}_F = 30c$ for the patch at complete fossilization (equation 7). Thus, $\mathrm{Re}/\mathrm{Re}_F = \varepsilon/\varepsilon_F$.

As shown in the HPD of Figure 5, the upper right quadrant represents patches that are fully turbulent because the normalized Froude and Reynolds numbers are both supercritical ($\varepsilon \geq \varepsilon_o \geq \varepsilon_F$). In oceans and lakes most patches are found in the active-fossil quadrant ($\varepsilon_o \geq \varepsilon \geq \varepsilon_F$), indicating that the largest scales are fossil and the smallest scales are active and overturning. Fully fossil turbulence patches ($\varepsilon_o \geq \varepsilon_F \geq \varepsilon$) are only found in laboratories. In the ocean, ambient internal wave motions supply energy to fossil turbulence patches, so their ε values remain larger than the fully fossil value $\varepsilon_F$.

The laboratory decay line with slope 1/3 shown in Figure 5 represents a series of stratified grid turbulence experiments carried out at UCSD by Charles Van Atta and his students that demonstrated the fossil turbulence phenomenon and confirmed the predicted universal constants, Gibson (1987, 1991d). The schematic line with arrows shows the evolution of a turbulence patch in the ocean starting from critical Fr and Re values where the patch becomes fully turbulent and growing to large scales and large Re values until the Froude number becomes critical and fossilization begins along the 1/3 slope decay line. Because fossil turbulence patches do not collapse as often assumed but retain their maximum Thorpe overturn scale, they retain information about their initial large values of viscous and scalar dissipation rates. These can be used in hydro-paleontology, such as the modeling of vertical diffusivities in oceans and lakes taking advantage of information about previous powerful turbulence and turbulent mixing events preserved by fossil turbulence remnants, Gibson (1991b).

## 4. FOSSIL TURBULENCE IN THE OCEAN

Active turbulence is extremely rare in the ocean because it is rapidly damped by buoyancy, Coriolis and viscous forces. Potential entropy, represented by the variance of temperature, salinity, density, chemical species, and biological species, is produced by the active turbulence stirring. The scalar fluctuations are gradually and irreversibly mixed later and elsewhere by active-fossil and fossil turbulence motions to produce the entropy of mixing. Most mixing and diffusion in the ocean is initiated by active turbulence, but is then completed by fossil turbulence and zombie turbulence processes (zombie turbulence is reactivated fossil turbulence). We show evidence in the following that fossil density turbulence patches extract ambient internal wave energy and re-radiate the energy vertically at the ambient stratification frequency in a vertically beamed zombie turbulence maser action where the fossil patches are "pumped" to metastable states by ambient internal wave energy and information about the submerged fossils and the ambient internal waves are radiated to the surface where it can be extracted by remote sensing. The mixing process spreads vertically to much larger volumes and to other ocean layers as fossil turbulence patches within the layers. Kinetic energy is radiated as fossil turbulence waves, re-absorbed from ambient internal waves, and dissipated by viscous forces. Fossil turbulence processes are complex, but are crucially involved in all aspects of mixing and diffusion in stratified and rotating fluids that become actively turbulent.

**4.1 Data from the Sand Island Municipal Outfall study**

During the first week of September 2002, an extensive study of the stratified mixing and diffusion processes near the Honolulu Sand Island municipal wastewater outfall in Mamala Bay, Hawaii was conducted as part of the Remote Anthropogenic Sensing Program (RASP) with Russian colleagues. Two microstructure profilers (MicroStructure Measurement System, MSS) were used to horizontally and vertically profile standard CTD parameters plus microstructure parameters (temperature gradient, velocity shear, micro-conductivity, turbidity and fluorescence). The MSS instruments were made by Sea & Sun Technology and are described by Prandke and Stips (1998). The objective of this project is to provide "sea truth" microstructure measurements to establish the location and hydrodynamic properties of the waste field plume emanating from the diffuser for comparison with the plume location and properties inferred by remote observations of surface wave effects from space satellites and helicopters. The details and results of this project will be the subject of future papers. However, some preliminary hydrodynamic phase diagram results from the vertical profiles will be discussed in the context of the present paper as a demonstration of how fossil turbulence and HPDs can be useful to interpret oceanic mixing and diffusion processes.

Sand Island outfall is a 1.98 m diameter pipe that extends 2.4 km offshore to ocean depths of 69 to 73 meters and continuously discharges about 3 to 4 $m^3 s^{-1}$ of waste water (sewage after advanced primary treatment) into the stratified ocean as a buoyant turbulent plume. Instead of discharging in one big jet from the end of the pipe, the outfall is designed with a 1,040 m long diffuser section with 282 small jets that cause maximum initial dilution with dense bottom seawater. The rising plume mixes with the ambient ocean water until the diluted effluent reaches a trapping depth below the surface where its density matches that of the stratified receiving water (Koh 1975). Normally ocean microstructure studies have highly uncertain initial conditions where it is not possible to study the evolution of the stratified turbulence from active to fossilized hydrodynamic states. This study is the first application of modern microstructure instrumentation in a wastewater outfall, so that we can study the growth of the buoyancy driven actively turbulent plume as the turbulence scale grows and fossilizes at the trapping depth. The wastewater outfall represents a field laboratory to test stratified oceanic turbulence, mixing, and diffusion processes with known and reproducible forcing and ambient hydrophysical conditions.

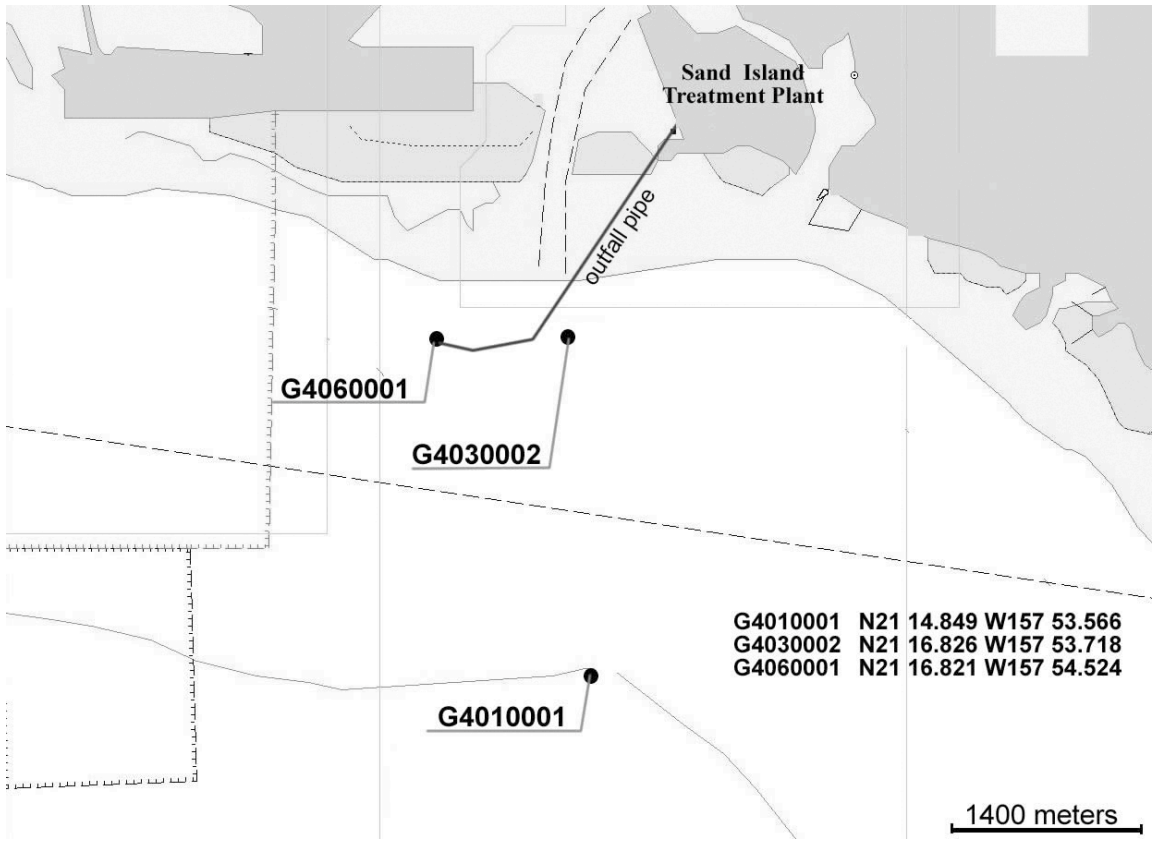

***Figure 6.*** A GPS map of the Sand Island sewage outfall area, and the stations used for sample HPD calculation (Fig. 9). Surface wave anomalies from internal waves radiated by the fossil turbulence wastefield were detected from space satellite images at distances up to 10 km south of the diffuser, and are correlated with microstructure patterns detected at depths above the wastefield by MSS sensors on a tow body.

Seven patches are selected from two vertical profiles (Figure 7 and 8) obtained at station G4010001 and G4060001. Station G4010001 is 3 kilometers south of the outfall diffuser port and G4060001 is directly over the end of the diffuser section (Figure 6). Equation 6 and 9 are used to estimate $\varepsilon_F$ and $\varepsilon_o$, where $L_{Tmax}$ in Equation 9 is the maximum Thorpe displacement scale of the patch (Thorpe 1977). N should be calculated from the ambient density gradient the turbulent patch is experiencing. In these sample calculations, the density gradient for N is computed from the monotonic Thorpe reordered density profile averaged between depths $1/2$ $L_{Tmax}$ above and below the patch. For patch A, N was estimated from the re-ordered density profile between 68.3 and 68.8 meters (Table 2). The normalized Froude number and normalized Reynolds number were then calculated as discussed in Section 3.

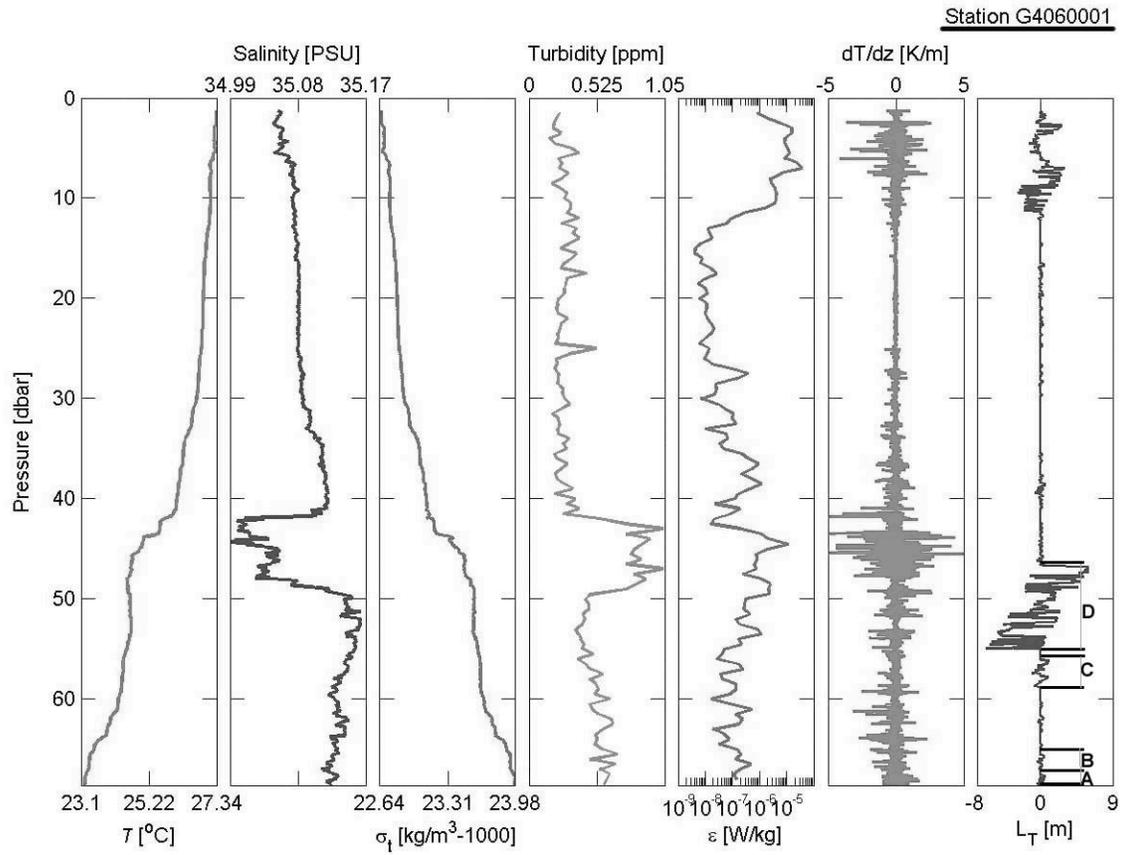

***Figure 7.*** Vertical MSS profile near the end of the diffuser, on 09/02/2002. From the left, temperature ($^o$C), salinity (PSU), density (kg/m$^3$ -1000), turbidity (ppm), dissipation rate (W/kg), temperature gradient dT/dz (K/m), and Thorpe displacement scale $L_T$ (m) are shown. The low salinity, high turbidity signature of the trapped wastefield is seen at 42-50 meters depth. Microstructure patches A, B, C, and D were identified for analysis from the Thorpe density displacement profile on the right.

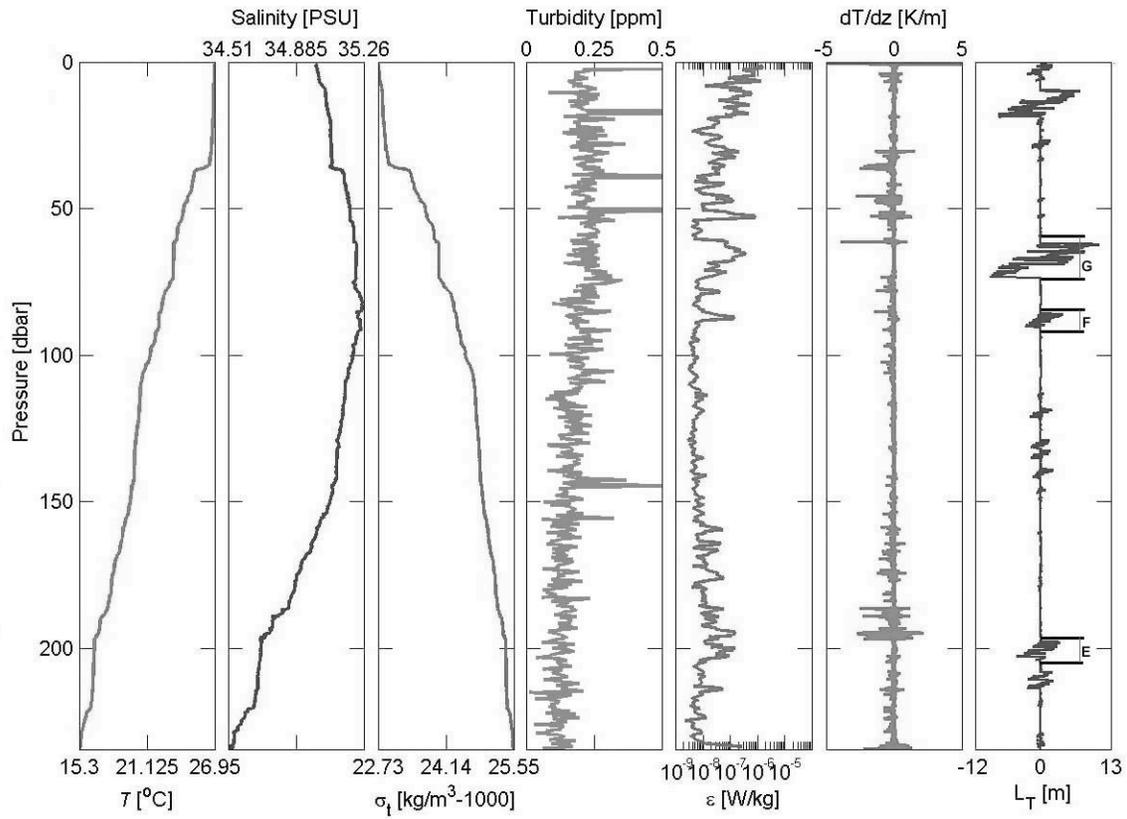

***Figure 8.*** Vertical profile from the station about 3 km south of the outfall diffuser. Microstructure patches E, F, and G were identified for analysis from the Thorpe displacement profile on the right.

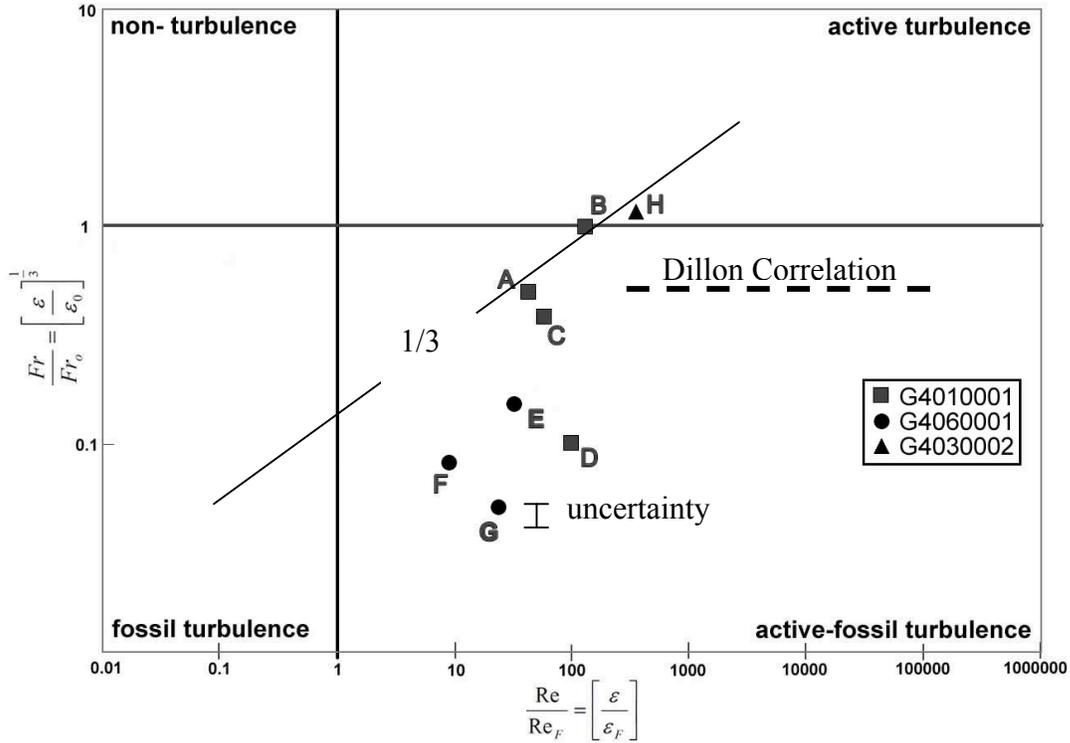

**Figure 9.** Hydrodynamic phase diagram (HPD) samples from two dropsonde stations and an ambient wind mixing station near the Sand Island Outfall, September 2, 2002. The points scatter widely, contrary to the "Dillon correlation" between $L_R$ and $L_T$.

Points A to D from Station G4060001, Figure 7, are directly over the end of the diffuser section with HPD points A and B just 2 and 4 m above the bottom where it is expected that the flow will be closest to active turbulent, as shown.

Points E to G from Station G4010001, Figure 8, are located 3 kilometers southeast of the last diffuser port of the outfall and show no turbidity but stronger fossilization of equally strong turbulence events from the slope 1/3 lines extrapolating to $Re_o/Re_F$. This typical of microstructure patches in the ocean interior, which are almost never found in their original fully turbulent state.

Point H from a station G4030002 just east of the diffuser was chosen to be representative of the surface wind mixing; the HPD point H is in the active-turbulence region of the diagram as expected. We show elsewhere that the surface turbulence patches above the large fossilized patches D and G produced by the trapped wastefield were radiated by fossil turbulence waves and were sufficiently more active than H that they could be detected by remote sensing methods.

| Patch | Depth min | Depth max | $L_T$ max | $\dfrac{Re}{Re_F} = \dfrac{\varepsilon}{\varepsilon_F}$ | $\dfrac{Fr}{Fr_0} = \left[\dfrac{\varepsilon}{\varepsilon_0}\right]^{\frac{1}{3}}$ | Station |
|---|---|---|---|---|---|---|
| A | 68 | 68.5 | 0.6 | 49.96 | 0.51 | G4060001 |
| B | 65.7 | 66.9 | 0.4 | 121.37 | 0.93 | G4060001 |
| C | 56.2 | 57.4 | 1.1 | 485.62 | 0.99 | G4060001 |
| D | 46.7 | 54.9 | 6.9 | 186.49 | 0.13 | G4060001 |
| E | 197.3 | 203.9 | 4.1 | 267.16 | 0.43 | G4010001 |
| F | 85.8 | 90.4 | 4.1 | 31.78 | 0.14 | G4010001 |
| G | 62.1 | 73.3 | 10.8 | 632.36 | 0.28 | G4010001 |
| H | 3.2 | 5.6 | 0.5 | 344.02 | 1.11 | G4030002 |

***Table 2:*** Values used to prepare the HPD shown in Figure 9.

      Length Scale ($L_T$) is the maximum Thorpe overturn scale of the patch.

      Depth (meters) is the location of the patch.

      ε is the dissipation rate of the patch measured by Microstructure Profilers

      $\varepsilon_o \approx 3L_T^2 N^3$ is the dissipation rate when the patch begins to fossilize.

      $\varepsilon_F = 30\nu N^2$ is the dissipation rate when the patch reaches complete fossilization.

      $Fr/Fr_o = \left(\varepsilon/\varepsilon_o\right)^{1/3}$ is the normalized Froude number.

      $Re/Re_F = \varepsilon/\varepsilon_F$ is the normalized Reynolds number.

The HPD points for patches A to D in Figure 7 are plotted on the hydrodynamic phase diagram shown in Figure 9. Wastewater ejected from the diffuser undergoes turbulent mixing. As the eddies grow larger and move away from the turbulent source, buoyancy forces become dominant and fossilize turbulent patches in the plume. Point A and B of Figure 9 are 0.5 m patches measured at depths 2 and 4 m above the end of the diffuser. Both are close to the completely active turbulence quadrant of the HPD, with $Fr/Fr_o$ almost equal to 1. The overturn length scales $L_{Tmax}$ of patches A, B, C, and D monotonically increase from 0.5 m to nearly 8 m as the turbulent plume rises to the trapping depth and the turbulence cascades from small scales to large as expected. Note that the largest patch C is strongly fossilized, showing that it has not collapsed due to buoyancy forces, contrary to the commonly accepted Gregg (1987) model. The small uncertainty shows there is no "Dillon correlation" constant ratio between $L_R$ and $L_T$ as assumed by non-fossil-turbulence models such as Dillon (1982, 1984), where $\varepsilon/\varepsilon_o \approx 1/3$.

For station G4010001 about 3 km south of the diffuser, Figure 7, the microstructure patches were all in the lower active-fossil quadrant, Figure 9, indicating fossilization at the largest scales but active overturning turbulence at smaller scales. Patch H was chosen in the surface wind mixed layer at a station about 300 m east of the diffuser to use for comparison with surface layer patches produced by radiation of fossil turbulence waves from below that might cause anomalous effects on the surface waves produced by the wind. Surface wave anomalies are detectable by remote sensing, and will be discussed elsewhere.

To summarize the work in progress, 1405 HPD points have now been computed from ambient N values computed from Thorpe re-ordered density microstructure patches with ε averaged over the patch. Of these 215 patches were in the active turbulence quadrant, 1185 were active-fossil, and only 5 were completely fossilized. The lack of completely fossilized patches suggests that fossil turbulence patches are extracting energy from the ambient internal wave and current motions of their surroundings. Many of the fully active turbulence patches were found just above the largest overturn fossils, suggesting they were triggered by the vertical internal wave radiation that appears to account for the remote sensing mechanism. Strong density gradients at the tops of fossil turbulence patches produce vorticity and turbulence when tilted by ambient internal waves, which then fossilize and radiate fossil turbulence internal waves nearly vertically. Turbulent patches form when the waves reach the weakly stratified surface layers, and these interfere with the formation of capillary waves by the wind. The term zombie turbulence, invented by Hide Yamazaki (personal communication 1990), aptly describes the process of fossil turbulence coming back to life.  The ambient stratification determines the frequency of the radiated internal waves and the vertical radiation, extracting ambient energy in a process similar to beamed maser or laser action.

Other ocean studies show evidence of vertical propagation of information about the existence of deep powerful turbulence sources.  Large Thorpe overturning scales are observed at depths up to 2 km directly above strong turbulent mixing zones for bottom currents passing through the Romanche fracture zone of the mid-Atlantic ridge, Ferron et al. (1998, Fig. 8).  Finnagan et al. (2002) report density overturning with indicated $L_{Tmax}$ values up to 50 meters (Fig. 3) at 860 m depth over the Hawaiian ridge at 2500 m, which we interpret as possible vertical radiation from agitated fossil turbulence (beamed zombie turbulence maser action) of strong turbulent mixing at the ridge and they interpret as breaking of baroclinic internal waves scattered horizontally from the Hawaiian ridge. Their estimates of dissipation rates from a Dillon (1982) correlation $L_R/L_{Trms} = 0.8$ without computation of HPDs or intermittency factors are likely to be substantial underestimates, Gibson (1991b).  For example, the dissipation rate at fossilization for the 50 m patch displayed in their Fig. 3 is $\varepsilon_o = 6\times10^{-7}$ m$^2$ s$^{-3}$ compared to $\varepsilon = 5\times10^{-10}$ m$^2$ s$^{-3}$ in their Fig. 4.

### 4.2 Hydrodynamic phase diagram summary

Figure 10 shows a summary of oceanic microstructure, Gibson (1996), classified according to the hydrodynamic state of the microstructure patches using an HPD.

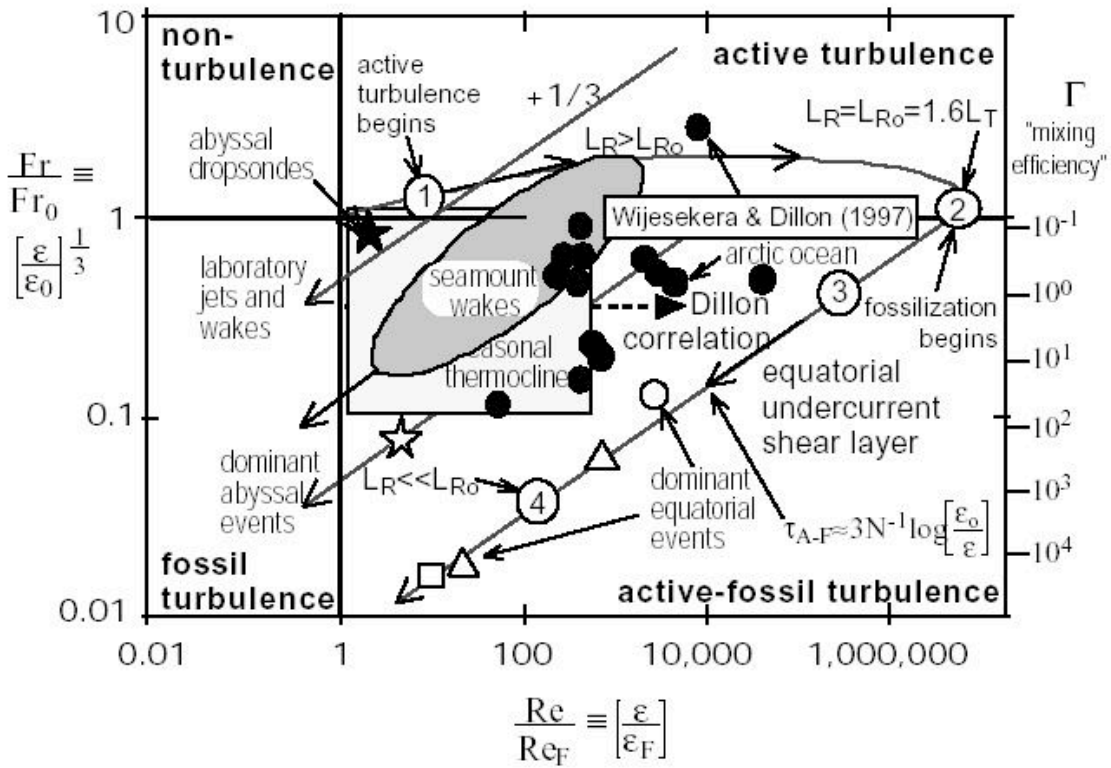

**Figure 10.** Hydrodynamic phase diagram for microstructure in the laboratory, seasonal thermocline, abyssal layers, seamount wakes, Arctic, and equatorial undercurrent (adapted from Gibson 1996).

1. *The Dillon (1982, 1984) correlation $L_R = L_{Trms}$ shown by the horizontal dashed arrow assumes seasonal thermocline patches are actively turbulent rather than active-fossil. The scatter is assumed to be statistical and the vertical shift is assumed to be due to errors in the Gibson (1980) universal constants. None of these assumptions are valid as shown by Figure 10.*

2. *Large equatorial patches with density overturns of twenty to thirty meters have been observed by Hebert et al. (1992) ○, Wijesekera and Dillon (1991) △, and Peters et al. (1994) □, indicating $\varepsilon_o$ values in previous actively turbulent states more than $10^6$ $\varepsilon_F$ and $10^3$-$10^5$ larger than measured values of $\varepsilon$ for these dominant equatorial events.*

3. *Abyssal dropsonde profiles ✱ Toole, Polzin and Schmitt (1994), are usually so sparse they fail to detect even the fossils of the dominant turbulence events, ☆ (Gibson 1982a).*

4. *The "mixing efficiency" $\Gamma \equiv (DCN^2/\varepsilon) = \varepsilon_o/13\varepsilon$ shown on the right (Oakey 1982) can be $10^4$ larger than 100% for fossilized patches in the deep ocean (see Gibson 1982b).*

5. *Measurements of powerful patches in the Arctic, ● Wijesekera and Dillon (1997) confirm the fact that large, actively-turbulent microstructure events exist in the ocean interior, even though they are rare.*

As shown in Figure 10, the laboratory studies have much smaller $Re_o$ to $Re_F$ ratios than most reported microstructure in ocean layers. Extensive measurements of microstructure were made during the MILE expedition, where simultaneous dropsonde and towed body measurements were possible. Since most microstructure patches found were relatively weak, Dillon (1984) proposed that possibly they were actively turbulent and that the universal constants had been incorrectly estimated in Gibson (1980). Dillon's correlation is shown by the horizontal arrow. When Dillon's correlation was proposed, few patches had been observed in their actively turbulent state except by towed bodies, despite laboratory confirmations of the Gibson (1980) universal constants by Stillinger et al. (1983) and Itsweire et al. (1986) in a stratified grid turbulence flow, as discussed by Gibson (1991d). However, measurements in the wake of Ampere Seamount, Gibson et al. (1994), showed that oceanic microstructure indeed begins in the actively turbulent quadrant for patches sampled over the crest of the seamount, and decays into the active-fossil quadrant for patches sampled downstream of the crest, as indicated by the oval region and arrow, confirming the universal constants of Gibson (1980) just as they are confirmed by Figure 9. From Figure 10 it follows that ocean microstructure is not likely to be found in its original actively turbulent state, but in some stage of fossilization. Wijesekera and Dillon (1997) reported only one fully active patch from their Arctic Sea data (dark circles) even though they were at high latitudes where Coriolis forces are large and the intermittency of dominant turbulence events is likely to be minimal.

### 4.3 Dark mixing paradox

Tom Dillon (personal communication) has suggested that the most important outstanding problem of physical oceanography today is the resolution of the "dark mixing paradox"; that is, unobserved mixing that must exist somewhere in the ocean to explain the fact that the ocean is mixed. The largest discrepancies between flux dissipation rate estimates of vertical diffusivities and those inferred from bulk flow models are in strongly stratified layers of the upper ocean such as the seasonal thermocline and the equatorial undercurrent, and deep in the main thermocline, Gibson (1990, 1991abcd, 1996).

The discrepancies disappear when the evidence of vast under-sampling provided by fossil turbulence is taken into account, and the extreme lognormal intermittency of the dissipation rates are used to estimate mean values, rather than using typical values that are closer to the mode rather than the mean. The mean to mode ratio of a lognormal random variable is $\exp(3\sigma^2/2)$, where $\sigma^2$ is the intermittency factor, or variance of the natural logarithm of $\varepsilon$ about the mean, of the probability distribution function. Baker and Gibson (1987) have shown that $\sigma^2$ for $\varepsilon$ and $\chi$ range from 3-7 in the ocean, showing that under-sampling errors are probably in the range 90 to 36,000 if the fossil turbulence evidence is ignored and no attempt is made to compensate for intermittency. Gibson (1991d) shows that when the observed lognormality of dissipation rates in the deep thermocline is taken into account, there is no discrepancy between the vertical diffusivity of temperature inferred using the Munk (1966) abyssal recipe, and that inferred from the Gregg (1977) deep Cox number measurements.

### 4.4 Fossil turbulence and turbulence effects on phytoplankton growth

Opposite effects of fossil turbulence and turbulence properties has been discovered on the growth rates of swimming and non-swimming phytoplankton species. It has been known for more than 60 years by marine biologists that phytoplankton growth is sensitive to turbulence, particularly red tide dinoflagellates whose growth is inhibited by turbulence. Allen (1938, 1942, 1943, 1946ab) observed that red tides in La Jolla bay were preceded by at least a two week period of sunny days and light winds. Red tides would not occur otherwise. Margalev (1978) suggested a "mandala" (a schematic representation of the cosmos in eastern art and religion) for phytoplankton survival where dinoflagellates, that swim, are favored when turbulence levels are low and diatoms, that can't swim, are favored when turbulence levels are high. Efforts to grow dinoflagellate species in the laboratory typically fail when the cultures are aerated or shaken vigorously for long periods.

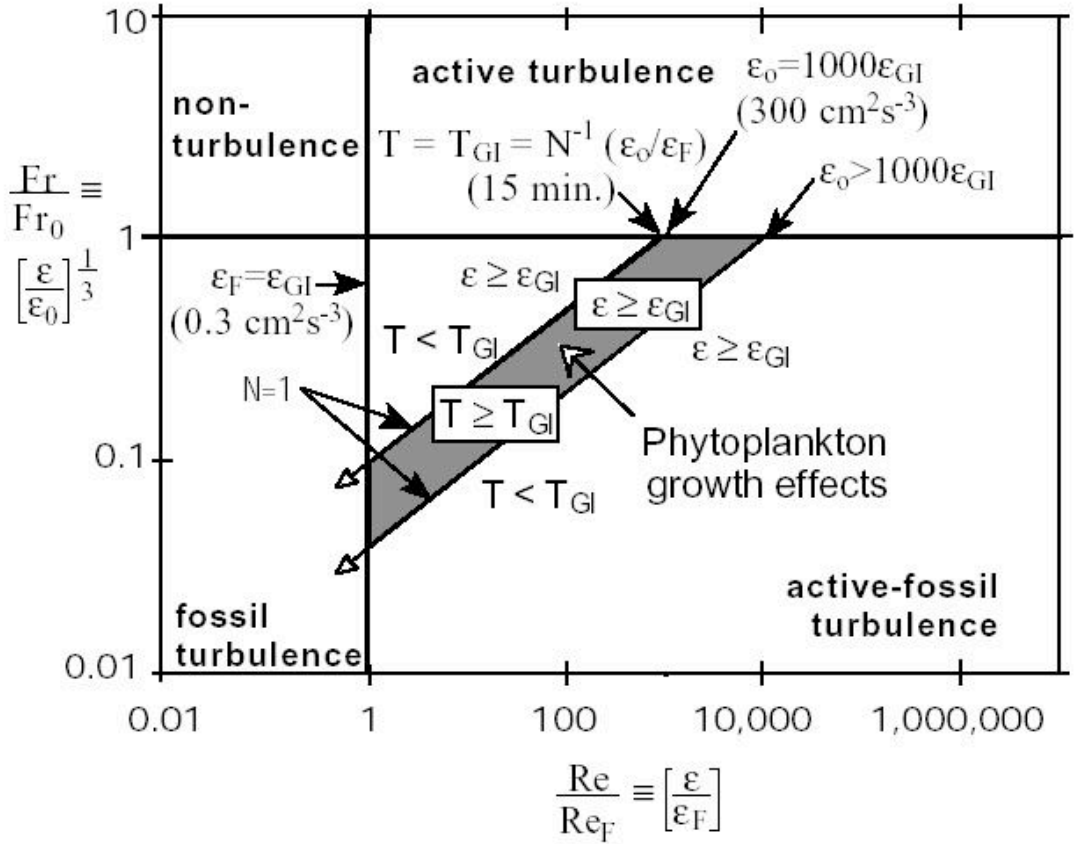

***Figure 11.*** Hydrodynamic phase diagram for phytoplankton growth effects based on the
G. polyedra laboratory results. The constraints $\varepsilon > \varepsilon_{GI}$ for time $T > T_{GI}$ are satisfied
only for N = 1 rad/s and turbulence at fossilization $\varepsilon_0$ values $\geq 1000 \, \varepsilon_{GI}$ (from
Gibson and Thomas 1995).

Using a Couette flow between two concentric cylinders to achieve a uniform rate-
of-strain $\gamma$ for growing phytoplankton cultures, compared to growth between identical
motionless control cylinders, Thomas and Gibson (1990) quantified the growth response
of the red tide dinoflagellate Gonyaulax *polyedra* Stein. It was found that the viscous
dissipation rate $\varepsilon \approx \nu \gamma^2$ for growth inhibition $\varepsilon_{GI}$ was about $2\text{-}3 \times 10^{-5}$ m$^2$s$^{-3}$. Other
dinoflagellate species tested also showed a negative response to continuous strain rates.
In the case of *Prorocentrum micans*, negative growth rate was observed only for high
light levels of 332-451 $\mu$E m$^{-2}$ s$^{-1}$ and not for lower values of 162-209 $\mu$E m$^{-2}$s$^{-1}$, Tynan,
Thomas and Gibson (1997).

Gibson and Thomas (1995) report the effects of turbulence intermittency on
dinoflagellate growth, based on laboratory studies and the field tests of Tynan (1993).
The remarkable result found for Gonyaulax *polyedra* was that continuous strain rates
with $\varepsilon > \varepsilon_{GI}$ have the same or greater negative effects on growth when applied for short
time periods T greater than a minimum $T_{GI}$ during a day for several days, where $T_{GI}$ was
in the range of only 5-15 minutes. Thus, intermittency of turbulence reduces the daily average ε threshold values for growth inhibition by factors of 100 or more. Vigorous stresses applied for short periods (<< 5 minutes) were ignored by the dinoflagellate cultures; for example, when the total culture was at least daily mixed together to determine an accurate average cell concentration during the ten day tests. Dissipation ε values to cause maximum bioluminescence are about $10^5$ $\varepsilon_{GI}$ for Gonyaulax *polyedra* and are non-fatal, based on viscous stress levels of τ ≈ 10 dynes/cm², Rohr et al. (1997). Tynan (1993) found a negative correlation of the dinoflagellate population (after a three day lag) with significant wave height and wave velocity, and the opposite response for the diatom population. Neither dinoflagellate nor diatom populations were strongly correlated with wind speed. Thomas, Tynan and Gibson (1997) report positive effects of laboratory turbulence on diatom growth, although the effects of intermittency on diatoms have not been tested in the laboratory.

To explain this response, Gibson and Thomas (1995) propose the possibility that phytoplankton have evolved the ability to recognize patterns of turbulence-fossil-turbulence in the surface layer of the sea. Evolution has adjusted their growth rates in response to these patterns to optimize their competitive advantages with respect to swimming ability. Diatoms require turbulence in the surface layer to bring them up to the light. When the surface layer is strongly stratified by sunny days and long periods without turbulence, dinoflagellates can swim toward the light and bloom if adequate nutrients are available, but diatoms will sink out. Breaking surface waves produce intermittent turbulence patches in the euphotic zone that have very different ε values as a function of time depending on whether the fluid is stratified or non-stratified. If the surface layer is not stratified the dissipation spectrum $k^2\phi_u$ is the universal Kolmogorov form with +1/3 slope and persistence times of only a few seconds corresponding to a few overturn times of the largest eddies. If the surface layer is stratified, however, the dissipation rate is preserved above ambient by the fossilization process. From Gibson (1980) the decay time T for fossil vorticity turbulence is about $N^{-1}(\varepsilon_o/\varepsilon_F)$. Breaking surface waves have very large ε values compared to those measured in the ocean interior which are generally much less than $\varepsilon_{GI}$ values indicated for any dinoflagellate species. Dissipation rates measured in the surf zone were found to be ε = 5-50×$10^{-5}$ m²s⁻³ by George et al. (1994), which could give ε > $\varepsilon_{GI}$ for periods T much longer than $T_{GI}$ if the surface layer is strongly stratified so that $\varepsilon_F > \varepsilon_{GI}$. However, the appearance of breaking surface waves in one location more than once a day for several days is a good indication that a sea change is in progress from a strongly stratified surface layer, favorable to dinoflagellates that can swim, to a turbulent mixed surface layer, favorable to diatoms that cannot. The model is illustrated by an example on a hydrodynamic phase diagram in Figure 11 using threshold values $\varepsilon_{GI}$ and $T_{GI}$ measured for dinoflagellate *Gonyaulax polyedra* Stein.

In Figure 11 it is assumed that $\varepsilon_{GI}$ = 3×$10^{-5}$ m²s⁻³, so that N ≈ 1 rad/s to give $\varepsilon_F$ = 3×$10^{-5}$ m²s⁻³. This is a strong stratification for the ocean, but could happen near the surface under red tide conditions. For a breaking wave patch to persist for 15 minutes (900 s) with this stratification requires $(\varepsilon_o/\varepsilon_F) \approx 1000$, so $\varepsilon_o \approx 3×10^{-2}$ m²s⁻³ or greater, exceeding the large surf zone values measured by George et al. (1994). A shaded zone of possible phytoplankton growth effects is shown in Figure 11. To the left of the shaded region, surface wave patches will not persist long enough to have an effect. To the right of the region, the surface wave breaking is so powerful that the turbulent jet might penetrate below the euphotic layer. Larger N values require larger $\varepsilon_o$ values, and both are physically unlikely. Smaller N values might not give patches that persist long enough with $\varepsilon > \varepsilon_{GI}$, although the question of how dissipation rates evolve with time in a fossil turbulence patch is an important unsolved problem. A semi-empirical estimate of the time $\tau_{A-F}$ for $\varepsilon_O$ to decay to a value $\varepsilon$ in the active-fossil quadrant is given by the expression in Figure 11.

The active-fossil-turbulence pattern recognition scenario of Gibson and Thomas (1995) is intended as a working hypothesis that explains the presently available data for the response of phytoplankton to turbulence and turbulence intermittency. Other scenarios may exist, but have so far not been put forth.

## 5  SUMMARY AND CONCLUSION

The theory of turbulence is discussed from first principles. The instability due to $\vec{v} \times \vec{\omega}$ leading to turbulence is described and illustrated in Figure 1. The same instability and force balances apply at all stages of the turbulence cascade from small scales to large, which is the basis of universal similarity theories of turbulence, turbulent mixing and fossil turbulence. A precise definition of turbulence requires that irrotational cascades of energy from large scales to small are non-turbulent because inertial-vortex forces $\vec{v} \times \vec{\omega}$ are zero. A universal small-to-large direction of the turbulence cascade is critical to an unambiguous signature of fossil turbulence. Irrotational, inviscid Kelvin-Helmholtz billows are not only physically impossible to form at large scales as the assumed large-scale initial turbulent eddies that cascade to smaller scales, but are non-turbulent by our definition.

Fossil turbulence theory is introduced and fossil turbulence is defined. Most of the observed microstructure patches in sampled layers of the ocean are found to be fossilized from their classification as active, active-fossil, and fossil turbulence using $Fr/Fr_O$ versus $Re/Re_F$ hydrodynamic phase diagrams using the fossil turbulence theory of Gibson (1980, 1986). Vertical diffusivities estimated for these layers without taking fossil turbulence effects into account should be considered unreliable.

Fossil turbulence is observed in most ocean microstructure patches, especially the largest patches that dominate vertical fluxes through deep interior layers. Microstructure measurements should particularly be examined for the hydrodynamic state of those patches that dominate estimates of the average dissipation rates for the oceanic layer. If such patches are strongly fossilized, as expected since this has been the case for all dropsonde data sets so far reported, then the turbulence processes of the layer are likely to be undersampled and the average dissipation rates and vertical fluxes substantially underestimated. Fossil turbulence processes affect a wide variety of oceanic and limnological phenomena. Evidence is presented that the two major ocean phytoplankton species have evolved growth strategies reflecting surface layer fossil turbulence processes that affect their competitive advantages with respect to different swimming abilities.

Laboratory quality oceanic microstructure measurements made in a municipal wastewater outfall are presented that clearly demonstrate that stratified turbulence cascades from small scales to large and fossilizes without collapse, contrary to the commonly accepted non-fossil-turbulence theories of Dillon (1982, 1984) and Gregg (1987) which should be abandoned. These and other microstructure measurements described in the paper support narrow definitions of turbulence and fossil turbulence, Gibson (1991a, 1996, 1999), as well as the universal constants of the Gibson (1980, 1986, 1987) fossil turbulence theory. A new mechanism of vertical energy, mixing, and information transport is identified from the remote sensing of a submerged fossil-turbulence waste-field. The mechanism involves extraction of ambient internal wave energy by the fossil-density-turbulence patches, which then beam internal wave energy vertically to the surface where it forms active turbulence patches that interfere with the formation of capillary waves by the wind and permit remote detection. The mechanism is similar to that of beamed masers. The fossils are pumped to metastable states (zombie turbulence) by the ambient internal wave motions and the stratification frequency of the ambient ocean, which is the frequency of fossil-vorticity-turbulence, from Gibson (1980), and prescribes a preferred vertical direction of re-radiation.

# 6　ACKNOWLEDGEMENTS

We gratefully acknowledge the scientific support and encouragement of Dr. R. Norris Keeler throughout the several years of planning leading to the Remote Anthropogenic Sensing Program and the Mamala Bay hydrodynamic phase diagram measurements. Fabian Wolk of Rockland Oceanographic Services Inc. and Hartmut Prandke of ISW Wassermesstechnik supplied the equipment, made the measurements and helped analyze the data. Financial support for this work was provided by NAVAIR contract NBCHF010272, through Directed Technologies, Inc. and logistical support in Hawaii was provided by Oceanit, Inc. We are also most grateful for a very helpful and constructive report provided by an anonymous reviewer.